\documentclass[a4paper,12pt]{article}
\usepackage{amsmath}
\usepackage{amsthm}
\usepackage{amsfonts}
\usepackage{mathrsfs}
\usepackage{graphicx}
\usepackage{hyperref}

\newcommand{\lR}{\mathrm{I\hspace{-0.7mm}R}}

\numberwithin{equation}{section}

\hypersetup{ pdftitle={Draft},
pdfauthor={Fiki T. Akbar, Bobby E. Gunara}, 
pdfkeywords={supersymmetry, local existence}, 
bookmarksnumbered, pdfstartview={FitH}, urlcolor=blue, }

\begin{document}
\pagestyle{plain}




\title{\LARGE\textbf{Higher Dimensional Curved Domain Walls on K\"ahler Surfaces}}

\author{{Fiki T. Akbar$^{\sharp}$, Bobby E. Gunara\footnote{Corresponding author}} $^{\sharp}$, \\ Flinn C. Radjabaycolle$^{\sharp,\flat}$,  Rio N. Wijaya$^{\sharp}$, \\ \\
$^{\sharp}$\textit{\small Theoretical Physics Laboratory}\\
\textit{\small Theoretical High Energy Physics and Instrumentation Research Group,}\\
\textit{\small Faculty of Mathematics and Natural Sciences,}\\
\textit{\small Institut Teknologi Bandung}\\
\textit{\small Jl. Ganesha no. 10 Bandung, Indonesia, 40132}\\
{\small and} \\
$^{\flat}$\textit{\small Departement of Physics}\\
\textit{\small Faculty of Mathematics and Natural Sciences,}\\
\textit{\small Cendrawasih University}\\
\textit{\small Jl. Kampwolker Kampus Uncen Baru Waena-Jayapura 99351, Indonesia} \\
\small email: ftakbar@fi.itb.ac.id, bobby@fi.itb.ac.id}

\date{}

\maketitle




\begin{abstract}
In this paper we study some aspects of curved BPS-like domain walls in higher dimensional gravity theory coupled to scalars where the scalars span a complex K\"ahler surface with scalar potential turned on. Assuming that a fake superpotential has a special form which depends on K\"ahler potential and a holomorphic function, we prove that BPS-like equations have a local unique solution. Then, we analyze the vacuum structure of the theory including their stability using dynamical system and their existence in ultraviolet-infrared regions using renormalization group flow.

\end{abstract}




\section{Introduction and Main Results}

Topological defects such as domain walls in  gravity-scalar systems have been studied over the last twenty years since they might provide some tools to explore the correspondence between  gravitational theory in anti-de Sitter  (AdS) spacetime and a conformal field theory (CFT) in flat Minkowskian spacetime (referred to as  AdS/CFT correspondence) both in the context of the context of supersymmetric theories and in the the context of non-supersymmetric theories. In the first situation, there has been a lot of study investigating these walls   in  five dimensional supergravity, see for example \cite{DWc, CDKV, CDL2002, CL2002}, while in the latter, some studies have been focused on higher dimensional models, see for example, \cite{GravStab, MA2012}. 

Apart from the above application, our interest here is to investigate the nature of curved domain walls of a higher dimensional gravity-scalar system with scalar potential turned on where the scalars are complex and span a two dimensional K\"ahler surface. These walls are described by an ansatz metric on which the codimension-one sliced spacetime has a cosmological constant, say $\Lambda_D$. In particular, we consider a model in which the scalar potential can be expressed in terms of a fake superpotential $\mathcal{W}(z, \bar{z})$ and its derivative (see eq. (\ref{CurvedPotential})) such that  the dynamics of the walls are governed by first order differential equations called Bogomol'nyi-Prasad-Sommerfeld (BPS)-like equations (or gradient flow equations). 

Our next step is to show that  these BPS-like equations have a unique local  solution. This can be achieved by rewriting BPS-like equations into the form of beta functions describing a renormalization group (RG) flow in the CFT picture. By taking a special case where $\mathcal{W}(z, \bar{z})$ depends on a real $C^{k+2}$ K\"ahler potential $K(z, \bar{z})$ for integer $k \ge 1$ and a holomorphic function $W(z)$ (see eq. (\ref{CurlWdef})), we can then construct a local $C^k$ Lipshitz function with respect to the scalars $(z, \bar{z})$ and further show that there exists a unique local  solution of BPS-like equations.

Second, our analysis on critical points of the fake superpotential $\mathcal{W}(z, \bar{z})$  shows that there are possibly three types of them, namely local minima, saddle points, and degenerate points. Each of them are related to vacuums (ground states) of the theory which can be viewed as critical points of the scalar potential. The ambient spacetimes at the vacuums become Ricci-flat or Einstein with negative cosmological constant. In the first case, the sliced spacetime should also Ricci-flat ($\Lambda_D = 0$) in order to evade ambient spacetime singularity, while in the latter case, $\Lambda_D$ could be arbitrary. 

If the model is "fake supersymmetric" introduced in \cite{GravStab, Townsend,  FNSS2004} with $\Lambda_D = 0$, then gravitational stability further requires that the sliced spacetime should be flat Minkowskian  and the ambient spacetimes simplify to maximally symmetric spacetimes, namely flat Minkowskian or maximally symmetric AdS spacetime. This conclusion is true  for general form of the metric of K\"ahler  surfaces and arbitrariness of the fake superpotential $\mathcal{W}(\phi)$. But, it becomes unworkable for $\Lambda_D \ne 0$ in general because so far,  the gravitational stability analysis using Witten-Nester energy functional, for example, in  \cite{FNSS2004}  does not cover this case. 

Finally, we conclude our analysis on gradient flow equations and RG flows at the ground states as follows. In Ricci-flat domain walls ( $\Lambda_D = 0$), the vacuums can be defined in the asymptotic region and can be categorized into ultraviolet (UV) vacuums in high energy level and infrared (IR) vacuums in low energy level. In the case of Einstein domain walls, such a setup, however, could not always be well-defined in general. A way to overcome this problem is to modify the assumption  that this theory can only be defined on a non-compact interval $I \subseteq \lR$ such that the IR regions can be defined in the limit of Inf($I$), while the UV regions are in the limit of Sup($I$). In addition, we get the same conclusions for both cases, namely in the UV regions, all types of ground states are allowed, whereas in the IR regions, only stable local minima are permitted.

The organization of this paper can be mentioned as follows. In Section \ref{DomainWallsIntrosection} we give a short discussion about domain walls of a higher dimensional gravity-scalar system where the scalars can be viewed as the complex coordinates of a  K\"ahler surface. We show the local existence and the uniqueness of BPS-like equations in Section \ref{LEUsection}. The discussion of vacuum structure of the theory is in Section \ref{CritSCurvedDMsection}.  We discuss the stability  of vacuums in the context of dynamical system using BPS-like equations and their classification using  beta functions in Section \ref{GFRGsection}. In the end, we apply the general analysis to two simple models. First, we consider a model with flat K\"ahler metric and quadratic holomorphic function $W(z)$, and discuss its relation to flat BPS domain walls of four dimensional $N=1$ supergravity coupled to a chiral multiplet. Second, we show that our setting appears naturally in the context of BPS domain walls of five dimensional $N=2$ supergravity coupled to a hypermultiplet.

\section{$(D+1)$-Dimensional Domain Walls on K\"ahler Surfaces}
\label{DomainWallsIntrosection}

In this section we shortly discuss a higher dimensional gravity theory coupled with a complex scalar where the scalar spans a K\"ahler surface explained below. Our setting here is inspired by our previous work on flat and curved domain walls of $N=1$ chiral supergravity in four dimensions \cite{GZA2007, GZ2009}.\\
\indent Let $\left(\mathcal{M}^{D+1}, g_{\mu\nu}\right)$ be a spacetime manifold in $D+1$ dimensions equipped with metric tensor $g_{\mu\nu}$ where $\mu,\nu = 0,1,\ldots,D$. In this paper, we are using the  signature $(-,+,\ldots,+)$. The action of the Einstein-Complex Scalar system in $D+1$ dimensions has the form
\begin{equation}
S = \int\: d^{D+1}x\:\sqrt{-g}\left(\frac{1}{2 \kappa^{2}}R - g_{z\bar{z}}\partial_{\mu}z\partial^{\mu}\bar{z} - V(z,\bar{z})\right)\:, \label{FakeSUGRAAction}
\end{equation}
where the above quantities can be mentioned as follows. The quantity $\kappa \equiv 1/M_P$ with $M_P$ is Planck mass, $g \equiv \det{(g_{\mu\nu})}$, and $R$ is a scalar Ricci of $\mathcal{M}^{D+1}$. The field $z$ is a complex scalar together with its conjugate $\bar{z}$  spanned a K\"ahler surfacewith metric $g_{z\bar{z}} = \partial^{2}K/\partial{z}\partial{\bar{z}}$ where $K(z,\bar{z})$ is  K\"ahler potential. In particular, we assume $g_{z\bar{z}} $ to be positive definite. Finally, the function $V(z,\bar{z})$ is a real scalar potential.

Then, the field equations of motions  of the theory consist of two equations, namely the Einstein field equation,
\begin{equation}
\frac{1}{\kappa^{2}}R_{\mu\nu} = g_{z\bar{z}}\left(\partial_{\mu}z\partial_{\nu}\bar{z} + \partial_{\nu}z\partial_{\mu}\bar{z} \right) + \frac{2}{D-1}g_{\mu\nu}V(z,\bar{z}) \:, \label{EinsteinFEq}
\end{equation}
and the scalar field equations of motions,
\begin{equation}
\frac{g_{z\bar{z}}}{\sqrt{-g}}\:\partial_{\mu}\left(\sqrt{-g}\partial^{\mu}z\right) + \partial_z g_{z\bar{z}} \partial_{\mu}z \partial^{\mu}z - \partial_{\bar{z}}V(z,\bar{z}) = 0\:, \label{ScalarFeq}
\end{equation}
together with their complex conjugate.

The domain walls solution can be constructed by considering the following ansatz of the spacetime metric
\begin{equation}
ds^{2} = a^{2}(u)\tilde{g}_{ab}(x^{a})dx^{a}dx^{b} + du^{2} \: ,  \label{CurvedDWmetric}
\end{equation}
where $x^{\mu} = (x^{a},u)$, $a,b = 0,1,\ldots,D-1$ are the local coordinates defined on $\mathcal{M}$ and $a(u)$ is the warp factor. The form of sliced metric $\tilde{g}_{ab}$ determine the type of domain walls.  In this paper, we consider the case of the curved domain walls on which the sliced metric $\tilde{g}_{ab}$ has to be Einstein  metric in $D$ dimensions satisfying
\begin{equation}
\tilde{R}_{ab} = \Lambda_{D} \: \tilde{g}_{ab}\:,
\end{equation}
where $\Lambda_{D}$ is a cosmological constant in $D$ dimensions known as Einstein-sliced domain walls. For $\Lambda_{D} = 0$, the sliced geometry is Ricci-flat and in general, this is still a curved spacetime. In this case we have Ricci-flat domain walls which generalize flat domain walls. For $\Lambda_{D}  > 0$ case, we have de Sitter (dS)-domain walls, whereas  anti-de Sitter (AdS)-domain walls  require $\Lambda_{D} < 0$.\\
\indent Next, one can compute the Ricci tensor related to the metric  (\ref{CurvedDWmetric}) which has the form
\begin{equation}
R_{\mu\nu} = -\tilde{g}_{ab}\delta^{a}_{\mu}\delta^{b}_{\nu}\left[(D-1)a^{'2} + aa''- \Lambda_{D} \right] - \delta^{D}_{\mu}\delta^{D}_{\nu} D \left(\frac{a''}{a}\right) \: ,
\end{equation}
and the Ricci scalar is given by
\begin{equation}
R = -D\left[(D-1)\left(\frac{a'}{a}\right)^{2} + 2 \frac{a''}{a}\: - \frac{\Lambda_{D}}{a^{2}}\right],
\end{equation}
where $a' \equiv  da/du$.

In the rest of the paper, we consider  domain wall solutions of (\ref{EinsteinFEq}) and (\ref{ScalarFeq}) for $u$-dependent scalar field. Taking $z \equiv z(u)$, the equations of motions (\ref{EinsteinFEq}) and (\ref{ScalarFeq}) can be reduced into second order ordinary differential equations with respect to $u$, namely
\begin{eqnarray}
\frac{a''}{a} + (D-1)\left(\frac{a'}{a}\right) - \frac{\Lambda_{D}}{a^{2}} + \frac{2\kappa^{2}}{(D-1)}V(z,\bar{z}) & = & 0\:,\\
\frac{a''}{a} + \frac{2\kappa^{2}}{D}g_{z\bar{z}}z'\bar{z}' + \frac{2\kappa^{2}}{D(D-1)}V(z,\bar{z}) & = & 0 \: , \label{CurvedWarpEq}
\end{eqnarray}
and
\begin{equation}
z''+\frac{D}{a}a'z' + g^{z\bar{z}}\partial_z g_{z\bar{z}}\: z' z' - g^{z\bar{z}}\partial_{\bar{z}}V(z,\bar{z}) = 0\:. \label{CurvedScalarEq}
\end{equation}
To solve (\ref{CurvedWarpEq}) -(\ref{CurvedScalarEq}), let us first define a set of first order flow equations
\begin{eqnarray}
\frac{a'}{a} & = & \pm \gamma(u)\mathcal{W}(z,\bar{z})\:, \label{CurvedProjectionEq} \\
z' & = & \mp \frac{(D-1)}{\kappa^{2}\gamma(u)}g^{z\bar{z}}\partial_{\bar{z}}\mathcal{W}\:, \label{CurvedBPSEq1} \\
\bar{z}' & = & \mp \frac{(D-1)}{\kappa^{2}\gamma(u)}g^{z\bar{z}}\partial_{z}\mathcal{W}\:,  \label{CurvedBPSEq2}
\end{eqnarray}
where $\mathcal{W} \equiv \mathcal{W}(z,\bar{z})$ is a real function (called fake superpotential) and the real valued phase factor $\gamma(u)$ is given by
\begin{equation}
	\gamma = \sqrt{1+\frac{\Lambda_{D}}{(D-1)a^{2}\mathcal{W}^2}} \ . \label{Gammadef}
\end{equation}
Equations (\ref{CurvedBPSEq1}) and (\ref{CurvedBPSEq2}) are known as gradient flow equations (or Bogomol'nyi-Prasad-Sommerfeld (BPS)-like equations).  Now, we can claim that the following statement is true. Any solution of equations (\ref{CurvedProjectionEq}), (\ref{CurvedBPSEq1}) and (\ref{CurvedBPSEq2}) are also the solution of equations (\ref{CurvedWarpEq}) and (\ref{CurvedScalarEq}) if  the following relation holds \cite{FNSS2004, DFGK2000}
\begin{equation}
\kappa^{2}V(z,\bar{z}) = \frac{(D-1)^{2}}{\kappa^{2}\gamma^{2}(u)}\: g^{z\bar{z}}\partial_{z}\mathcal{W}\:\partial_{\bar{z}}\mathcal{W} - \frac{D(D-1)}{2}\mathcal{W}^{2}\:. \label{CurvedPotential}
\end{equation}

Furthermore, the above setup implies the existence of a Hamiltonian constraint which has been studied in four dimensional models, see for example, \cite{Ferrara:1997tw, Ceresole:2007wx}. This can be seen as follows. The Lagrangian defined in (\ref{FakeSUGRAAction}) reduces to the effective Lagrangian
\begin{equation}
{\mathcal L} = a^D  \frac{D}{2\kappa^2}   \left[ (D-1) \left( \frac{a '}{a}  \right)^2  + \frac{\Lambda_D}{a^2}\right]  - a^D  \Big[ g_{z\bar{z}}  z' \bar{z}'  + V \Big] ~ ,\label{EffectiveL}  
\end{equation}
since on the interval $I$ we have
\begin{equation}
\int_{u_i}^{u_s}  a^D{''} du = a^D{'}(u_s)  - a^D{'} (u_i) \in \lR ~,
\end{equation}
where $u_i = {\mathrm{Inf}}(I)$ and $u_s = {\mathrm{Sup}}(I)$. Then, there exists a constant $E \in \lR$ such that
\begin{equation}
E = a^D  \frac{D}{2\kappa^2}   \left[ (D-1) \left( \frac{a '}{a}  \right)^2  - \frac{\Lambda_D}{a^2}\right]  + a^D  \Big[ g_{z\bar{z}}  z' \bar{z}'  + V \Big] ~ ,\label{Hconstraints}  
\end{equation}
which is the Hamiltonian constraint in higher dimension.
%



\section{Local Existence and Uniqueness}
\label{LEUsection}

In this section, we prove the local existence and the uniqueness of  (\ref{CurvedProjectionEq}) and the gradient flow  equations (\ref{CurvedBPSEq1}) and (\ref{CurvedBPSEq2}). Our starting point is to rewrite equations (\ref{CurvedProjectionEq}), (\ref{CurvedBPSEq1}) and (\ref{CurvedBPSEq2})  into the form
\begin{eqnarray}
a \frac{dz}{da} & = & -\frac{D-1}{\kappa^{2}\gamma^{2}}g^{z\bar{z}}\frac{\partial_{\bar{z}}\mathcal{W}}{\mathcal{W}}\:, \nonumber \\
a \frac{d\bar{z}}{da} & = & -\frac{D-1}{\kappa^{2}\gamma^{2}}g^{z\bar{z}}\frac{\partial_{z}\mathcal{W}}{\mathcal{W}}\:. \label{Betaeq}
\end{eqnarray}
In the context of Gravity/CFT correspondence,   (\ref{Betaeq}) can be thought of the beta functions which gives a relevant renormalization group (RG) flow of the scalar fields \cite{DWc}.  We will show that the local existence and the uniqueness of equations (\ref{Betaeq}) imply the local existence and the uniqueness of equationsof original equations, (\ref{CurvedProjectionEq}), (\ref{CurvedBPSEq1}) and (\ref{CurvedBPSEq2}). 

Let us write equations (\ref{Betaeq}) into an integral form
\begin{equation}
\mathbf{\Phi}(a)  = \mathbf{\Phi}_{0} - \int_{a_{0}}^{a}\: \mathbf{\mathcal{F}}(s,\mathbf{\Phi}(s))\:ds\:,
\end{equation}    
with
\begin{equation}
\mathbf{\Phi} = \left(\begin{array}{c} z \\ \bar{z} \end{array}\right)\:, \qquad 
\mathbf{\mathcal{F}}(a, \mathbf{\Phi}) = \frac{(D-1)^2}{\kappa^{2}}\frac{g^{z\bar{z}}a\mathcal{W}}{(D-1)a^{2}\mathcal{W}^{2} + \Lambda_{D}} \left(\begin{array}{c}\partial_{\bar{z}}\mathcal{W} \\ \partial_{z}\mathcal{W} \end{array}\right)\:,
\end{equation}
where we already used the definition of $\gamma$ in equation (\ref{Gammadef}).\\
 \indent For the rest of this paper, we consider  a particular case where the form of $\mathcal{W}(z,\bar{z})$ is given by 
\begin{equation}
\mathcal{W}(z,\bar{z}) \equiv e^{\kappa^{2}K/2} |W(z)|  \: , \label{CurlWdef}
\end{equation}
where $W(z)$  is a holomorphic function\footnote{This form of $\mathcal{W}(z,\bar{z})$ appears in the context of BPS domain walls of four dimensional $N=1$ supergravity coupled to chiral multilplets, see for example, \cite{GZA2007, GZ2009}. }. Assuming that the K\"ahler potential $K(z,\bar{z})$ is a real $C^{k+2}$ function for integer $k\geq 1$. Then by definition, the K\"ahler metric is a $C^{k}$ function.
Since the superpotential $W(z)$ is a holomorphic function, then it is analytic, thus a $C^{\infty}$ function. So, from equation (\ref{CurlWdef}), the function $\mathcal{W}$ is a function of class $C^{k+2}$.

Furthermore, the function $\mathcal{F}(a,\mathbf{\Phi})$ is a $C^{k}$ function with respect to $\mathbf{\Phi}$. Since $k \geq 1$, then $\mathcal{F}(a,\mathbf{\Phi})$ is a locally Lipshitz function respect to second argument, hence for a real constant $M$, there is exist a constant $L_{M}$ which depend on $M$ such that,
\begin{equation}
\left\|\mathcal{F}(a,\mathbf{\Phi}) - \mathcal{F}(a,\tilde{\mathbf{\Phi}})\right\| \leq L_{M} \left\|\mathbf{\Phi} - \tilde{\mathbf{\Phi}}\right\|\:,
\end{equation}
whenever $\|\mathbf{\Phi}\|,\|\tilde{\mathbf{\Phi}}\| \leq M$.

Define an operator
\begin{equation}
{\mathcal K}(\mathbf{\Phi})(a) =  \mathbf{\Phi}_{0} - \int_{a_{0}}^{a}\: \mathbf{\mathcal{F}}(s,\mathbf{\Phi}(s))\:ds\:.
\end{equation}
The local existence is established by showing that the operator ${\mathcal K}$ is a contraction mapping. Let $J = [a_{0},a_{0}+\delta]$ is an interval for a small $\delta$ and $U$ is an open set on one dimensional K\"ahler manifold. Also let,
\begin{equation}
X = \left\{\mathbf{\Phi} \in C(J,U) : \mathbf{\Phi}(a_{0}) = \mathbf{\Phi}_{0},\:\|\mathbf{\Phi}\| \leq M \right\}\:.
\end{equation}
Defining a norm on $X$ as
\begin{equation}
\|\mathbf{\Phi}\|_{X} = \sup_{a\in J}\:\|\mathbf{\Phi}(a)\| \: ,
\end{equation}
 we have then the estimate
\begin{eqnarray}
\| {\mathcal K}(\mathbf{\Phi})\|_{X} & \leq &  \|\mathbf{\Phi}_{0}\|_{X} + \sup_{a\in J} \int_{a_{0}}^{a}\: \|\mathbf{\mathcal{F}}(s,\mathbf{\Phi}(s))\|\:ds \nonumber\\
& \leq & \|\mathbf{\Phi}_{0}\|_{X} + \sup_{a \in J} \left(\|\mathbf{\mathcal{F}}(0)\| + L_{M}\|\mathbf{\Phi}\|\right)s\nonumber\\
& \leq & \|\mathbf{\Phi}_{0}\|_{X} + \left(a_{0}+\delta\right)\left(\|\mathbf{\mathcal{F}}(0)\| + M L_{M}\right)\:.
\end{eqnarray}
Taking
\begin{equation}
\delta \leq \min\left(\frac{1}{L_{M}},\frac{1}{\|\mathbf{\mathcal{F}}(0)\| + M L_{M}}\right) - a_{0} \: , \label{DeltaCondition}
\end{equation}
we see that $ {\mathcal K} (\mathbf{\Phi})$ map $X$ into itself. Thus, we have the following estimate
\begin{eqnarray}
\| {\mathcal K} (\mathbf{\Phi}) - {\mathcal K}(\tilde{\mathbf{\Phi}})\|_{X} & \leq & \sup_{a\in J} \int_{a_{0}}^{a}\: \|\mathbf{\mathcal{F}}(s,\mathbf{\Phi}(s))-\mathbf{\mathcal{F}}(s,\tilde{\mathbf{\Phi}}(s))\|\:ds \nonumber\\
& \leq & \sup_{a\in J}\:\sup_{a_{0}\leq s \leq a} s\:\|\mathbf{\mathcal{F}}(s,\mathbf{\Phi}(s))-\mathbf{\mathcal{F}}(s,\tilde{\mathbf{\Phi}}(s))\| \nonumber\\
& \leq & L_{M}(a_{0}+\delta)\|\mathbf{\Phi} - \tilde{\mathbf{\Phi}}\|_{X}\:.
\end{eqnarray} 
Since $\delta$ satisfies  (\ref{DeltaCondition}), then ${\mathcal K}$ is a contraction mapping on $X$. By contraction mapping principle, for each initial value, there exists a unique local solution of differential equation (\ref{Betaeq}). In addition, since $\mathcal{F}(a,\mathbf{\Phi})$ is a $C^{k}$-function respect to $\mathbf{\Phi}$, it follows that the local solution of (\ref{Betaeq}) is a $C^{k+1}$-function.

Using the local solution of (\ref{Betaeq}), the projection equation (\ref{CurvedProjectionEq}) becomes an autonomous equation of $a$,
\begin{equation}
\frac{a'}{a}  =  \gamma(a)\mathcal{W}(\mathbf{\Phi}(a))\:. \label{ProjectionEqAuto}
\end{equation}    
Since $\mathcal{W}$ is  a $C^{k+2}$ function and  using similar way as above, then we have a unique local solution of equation (\ref{ProjectionEqAuto}). Finally, using the composition  
\begin{equation}
z(u) = (z \circ a)(u) \: ,
\end{equation}
 the differential equations (\ref{CurvedBPSEq1}) and (\ref{CurvedBPSEq2}) admit a unique local solution.

\section{Vacuum Structure of The Theory }
\label{CritSCurvedDMsection}

In this section, we investigate some aspects of critical points of scalar potential $V(z,\bar{z})$ (\ref{CurvedPotential}), which is referred to as ground states or vacua of the theory, and its relation to  critical points of the fake superpotential $\mathcal{W}(z,\bar{z})$. In other words, if  $\mathbf{\Phi}_{c} \equiv (z_{c},\bar{z}_{c})$ is a critical point of $\mathcal{W} (z,\bar{z})$, then
\begin{equation}
 \partial_z \mathcal{W} (\mathbf{\Phi}_{c}) = \frac{ \mathcal{W} (\mathbf{\Phi}_{c}) }{2 W({z_c})}  \nabla_{z}W(\mathbf{\Phi}_{c})  = 0 \ ,
 \label{criticalW}
\end{equation}
where $ \nabla_{z}W(\mathbf{\Phi}_{c}) \equiv \frac{dW}{dz}(z_c) + \kappa^2 K_{z}(\mathbf{\Phi}_{c}) W({z_c})$.
\subsection{Critical Points of $\mathcal{W}(z,\bar{z})$}
\label{CPcurlW}


For the case at hand, namely   (\ref{CurlWdef}), we can write down the eigenvalues and the determinant of Hessian matrix of $\mathcal{W} (z,\bar{z})$ evaluated at $\mathbf{\Phi}_{c}$ as
\begin{eqnarray}
\lambda^{\mathcal{W}}_{\pm} & = & \kappa^{2}g_{z\bar{z}}(\mathbf{\Phi}_{c})\mathcal{W}(\mathbf{\Phi}_{c}) \pm  2|\partial_z^2 \mathcal{W}(\mathbf{\Phi}_{c})| \:,\label{eigenW} \\
\det H_{\mathcal{W}} & = & \kappa^{4} g_{z\bar{z}}^2(\mathbf{\Phi}_{c})\mathcal{W}^2(\mathbf{\Phi}_{c}) - 4 |\partial^2_z\mathcal{W}(\mathbf{\Phi}_{c})|^2 \:, \label{detW}
\end{eqnarray}
where
\begin{equation}
\partial^2_z\mathcal{W}(\mathbf{\Phi}_{c}) = \frac{\kappa^{2}\mathcal{W}(\mathbf{\Phi}_{c})}{2W(z_c)}\left[K_{zz}(\mathbf{\Phi}_{c})W(\mathbf{\Phi}_{c}) + K_z(\mathbf{\Phi}_{c})\frac{dW}{dz}(z_{c}) + \frac{1}{\kappa^{2}}\frac{d^2W}{dz^2}(z_{c})\right]\:.
\end{equation}  
The point $\mathbf{\Phi}_{c}$ is said to be a local minimum if all eigenvalues in (\ref{eigenW}) are positive, hence
\begin{equation}
\frac{\kappa^{2}}{2} g_{z\bar{z}}(\mathbf{\Phi}_{c}) \mathcal{W}(\mathbf{\Phi}_{c}) > |\partial^2_z\mathcal{W}(\mathbf{\Phi}_{c})| \:,
\label{minW}
\end{equation}
while a saddle point requires
\begin{equation}
\frac{\kappa^{2}}{2} g_{z\bar{z}}(\mathbf{\Phi}_{c}) \mathcal{W}(\mathbf{\Phi}_{c}) < |\partial^2_z\mathcal{W}(\mathbf{\Phi}_{c})| \:,
\label{saddleW}
\end{equation}
Moreover, $\mathbf{\Phi}_{c}$ is a degenerate  critical point if
\begin{equation}
\frac{\kappa^{2}}{2} g_{z\bar{z}}(\mathbf{\Phi}_{c}) \mathcal{W}(\mathbf{\Phi}_{c}) = |\partial^2_z\mathcal{W}(\mathbf{\Phi}_{c})| \:.
\end{equation}
In general, there is no local maximum point in the model.

\subsection{Vacuums}
\label{SCurvedDMsection}


Now, we can discuss the vacuum structure of the theory which can be viewed as the critical points of the scalar potential  $V(z,\bar{z})$.  The results here generalize our previous results of flat and curved BPS domain walls  in four dimensions \cite{GZA2007, GZ2009}.

The eigenvalues and the determinant of the Hessian matrix of $V(z,\bar{z})$ evaluated at  $\mathbf{\Phi}_{c}$ can be written down as 
\begin{eqnarray}
\lambda^V_{\pm} & = & \frac{2}{\kappa^{4}\gamma^{2}}(D-1)^2 g^{z\bar{z}}\left(\mathbf{\Phi}_{c}\right)|\partial^2_z\mathcal{W}\left(\mathbf{\Phi}_{c}\right)|^2 + (D-1)\left[\frac{D-1}{2\gamma^{2}}-D\right]g_{z\bar{z}}\left(\mathbf{\Phi}_{c}\right)\mathcal{W}^2\left(\mathbf{\Phi}_{c}\right) \nonumber\\
& &\pm \frac{2}{\kappa^{2}}(D-1)\left[D-\frac{D-1}{\gamma^{2}}\right]\mathcal{W}\left(\mathbf{\Phi}_{c}\right)|\partial^2_z\mathcal{W}\left(\mathbf{\Phi}_{c}\right)|\:,\label{CurvedEigenV} \\
\det H_{V} & = & \frac{4\left(g^{z\bar{z}}\left(\mathbf{\Phi}_{c}\right)\right)^{2}}{\kappa^{8}\gamma^{4}}(D-1)^4|\partial^2_z\mathcal{W}\left(\mathbf{\Phi}_{c}\right)|^4 + \frac{1}{4}(D-1)^2\left[\frac{D-1}{2\gamma^{2}}- D \right]^{2} g_{z\bar{z}}^2\left(\mathbf{\Phi}_{c}\right) \mathcal{W}^4\left(\mathbf{\Phi}_{c}\right) \nonumber \\
& & -\frac{2}{\kappa^{4}}(D-1)^2\left[\left(\frac{D-1}{\gamma^{2}}-D\right)^{2} + D^2\right] \mathcal{W}^2\left(\mathbf{\Phi}_{c}\right)|\partial^2_z\mathcal{W}\left(\mathbf{\Phi}_{c}\right)|^2 \:. \label{CurvedDetV}
\end{eqnarray}
Our analysis  on  (\ref{CurvedEigenV})  and (\ref{CurvedDetV}) concludes that a local minimum of the scalar potential (or a stable vacuum) exists if
\begin{equation}
|\partial^2_z\mathcal{W}\left(\mathbf{\Phi}_{c}\right)| > \frac{\kappa^{2}g_{z\bar{z}}\left(\mathbf{\Phi}_{c}\right) \mathcal{W}\left(\mathbf{\Phi}_{c}\right)}{2(D-1)} \left[ D\left(2\gamma^2 -1\right)+1 \right] \ ,
\label{CurvedMinV}
\end{equation}
whereas a local maximum (or an unstable vacuum) requires
\begin{equation}
|\partial^2_z\mathcal{W}\left(\mathbf{\Phi}_{c}\right)| < \frac{\kappa^{2}}{2}g_{z\bar{z}}\left(\mathbf{\Phi}_{c}\right)\mathcal{W}\left(\mathbf{\Phi}_{c}\right)\:.
\label{CurvedMaxV}
\end{equation}
The existence of a saddle point (or a saddle vacuum) demands
\begin{equation}
\frac{\kappa^{2}}{2}g_{z\bar{z}}\left(\mathbf{\Phi}_{c}\right)\mathcal{W}\left(\mathbf{\Phi}_{c}\right) < |\partial^2_z\mathcal{W}\left(\mathbf{\Phi}_{c}\right)| < \frac{\kappa^{2}g_{z\bar{z}}\left(\mathbf{\Phi}_{c}\right) \mathcal{W}\left(\mathbf{\Phi}_{c}\right)}{2(D-1)}   \left[ D\left(2\gamma^2 -1\right)+1 \right]  \ ,
\label{CurvedSaddleV}
\end{equation}
while  a critical point is said to be degenerate if
\begin{eqnarray}
|\partial^2_z\mathcal{W}\left(\mathbf{\Phi}_{c}\right)| & = & \frac{\kappa^{2}g_{z\bar{z}}\left(\mathbf{\Phi}_{c}\right) \mathcal{W}\left(\mathbf{\Phi}_{c}\right)}{2(D-1)}  \left[ D\left(2\gamma^2 -1\right)+1 \right] \: , \label{CurvedDegenerateV1}\\ \nonumber \\
|\partial^2_z\mathcal{W}\left(\mathbf{\Phi}_{c}\right)| & = & \frac{\kappa^{2}}{2}g_{z\bar{z}}\left(\mathbf{\Phi}_{c}\right)\mathcal{W}\left(\mathbf{\Phi}_{c}\right)\:.
\label{CurvedDegenerateV2}
\end{eqnarray}
From equations (\ref{minW}), (\ref{saddleW}), (\ref{CurvedMaxV}) and (\ref{CurvedSaddleV}), any local minima of $\mathcal{W}(z,\bar{z})$ are mapped into local maxima of the scalar potential $V(z,\bar{z})$ and  saddle point of $\mathcal{W}(z,\bar{z})$ are mapped into saddles or local minima of the scalar potential $V(z,\bar{z})$. Moreover the equation (\ref{CurvedDegenerateV2}) comes naturally from the fact that degenerate critical points of $\mathcal{W}$ are mapped into degenerate points of $V(z,\bar{z})$. Since both Hessian matrix $H_{\mathcal{W}}$ and $H_V$ evaluated at the points vanish, these special points are referred to as \emph{intrinsic} degenerate critical points of $V(z,\bar{z})$.

\subsection{Ambient Spacetimes At Vacuums}
\label{Spacetatvacua}

Since the gradient of $\mathcal{W}(z,\bar{z})$  vanishes at  $\mathbf{\Phi}_{c}$, the ambient spacetime $\mathcal{M}^{D+1}$ becomes Einstein with negative scalar curvature  whose  Ricci tensor  is given by
\begin{equation}
R_{\mu\nu} = -D \ \mathcal{W}\left(\mathbf{\Phi}_{c}\right)  g_{\mu\nu} \ ,
\end{equation}
while the warped factor $a(u)$ has the form
\begin{equation}
a(u) = \frac{1}{2}\left[\left(a_{0}+\sqrt{a_{0}^2 + \frac{\Lambda_{D}}{(D-1)\mathcal{W}^2(\mathbf{\Phi}_{c})}}\right)e^{\mathcal{W}(\mathbf{\Phi}_{c}) u} - \frac{\Lambda_{D}}{(D-1)\mathcal{W}(\mathbf{\Phi}_{c})^2}\frac{e^{-\mathcal{W}(\mathbf{\Phi}_{c}) u}}{\left(a_{0}+\sqrt{a_{0}^2 + \frac{\Lambda_{D}}{(D-1)\mathcal{W}^2(\mathbf{\Phi}_{c})}}\right)}\right]  \ , \label{ScaleNearCP}
\end{equation}
 with $a_{0}$ is a non-zero real constant. So, there are two possibilities of the ambient spacetimes at vacuums, namely the Ricci-flat spacetimes  for $\mathcal{W}\left(\mathbf{\Phi}_{c}\right) =0$ and $AdS_{D+1}$ spacetimes for $\mathcal{W}\left(\mathbf{\Phi}_{c}\right) \neq 0$. 

 In the first case, it is necessary that the sliced geometry should also be  Ricci-flat, namely $\Lambda_D =0$,  in order to evade the singularity of the warped factor $a(u)$. Ricci-flatness further implies that the vacuums are determined  by the critical point of the holomorphic superpotential
\begin{equation}
\frac{dW}{dz}(z_c) = 0 \ .
\end{equation}
since $W(z_c) = 0$. Particularly, if we  consider particularly  a "fake supersymmetric" model where one can introduce a set of fake supersymmetric transformations of a gravitino and a spin-$\frac{1}{2}$ fermion together with  Witten-Nester energy functional \cite{FNSS2004, Townsend}, then we have a case as follows.  In order to have "fake supersymmetric" vacuums, both  ambient and sliced  spacetimes become  flat Minkowskian.  Such a configuration is gravitationally stable since the Witten-Nester energy vanishes.

In $AdS_{D+1}$ spacetimes, we must have non-zero  $\mathcal{W}(\mathbf{\Phi}_{c})$ where $\mathbf{\Phi}_{c}$ is the solution of  the following nonholomorphic equation 
\begin{equation}
\nabla_{z}W(\mathbf{\Phi}_{c}) \equiv \frac{dW}{dz} (z_c) + \kappa^2 K_{z} (\mathbf{\Phi}_{c}) W (z_c) =0 \ . \label{criticalW1}
\end{equation}
For the case at hand, we might have several possibilities as follows. The sliced spacetime could be de Sitter ($\Lambda_D  > 0$), anti-de Sitter ($\Lambda_D < 0$), and Ricci-flat ($\Lambda_D =0$).  In addition, we could have a situation where the sliced cosmological constant $\Lambda_D$ is set to be $\Lambda_D = c \mathcal{W}^2\left(\mathbf{\Phi}_{c}\right)$ where $c$ is a non-zero real constant. In the latter case, as  $\mathcal{W}(\mathbf{\Phi}_{c}) \to 0$, we would regain the preceding Ricci-flat case. 

It is important to mention a remark as follows. For $\Lambda_D = 0$,   if the model is fake supersymmetric, then  the sliced spacetime simplifies to  flat Minkowskian and  the ambient spacetime becomes a maximally AdS spacetime. Such a configuration  is gravitationally stable. In the case of $\Lambda_D \ne 0$, it is however still unclear in general to check the gravitational stability of  (\ref{CurvedProjectionEq})-(\ref{CurvedBPSEq2}) with scalar potential (\ref{CurvedPotential}) because so far,  the conventional method using Witten-Nester energy functional, for example \cite{FNSS2004}, does not cover this case. 

\section{ Gradient Flows vs  RG Flows}
\label{GFRGsection}

In this section, we complete our analysis on the nature of vacuums using the gradient flow equations (\ref{CurvedBPSEq1})- (\ref{CurvedBPSEq2}) and  in context of the field theory using RG flow equations (\ref{Betaeq}) where the warped factor $a(u)$ can be viewed as an energy scale. In the case of Ricci-flat domain walls ($\Lambda_{D}=0$) the vacuum structure is simpler than Einstein domain walls ($\Lambda_{D} \ne 0$) since we do not need an additional analysis on the warped factor $a(u)$ as we will see below.

Let us start by considering the Ricci-flat domain walls.  The vacuums here are in the asymptotic region, \textit{i.e.} $u = \pm \infty$ and can be  classified into ultraviolet (UV) vacuums for $a \to +\infty$ and  infrared (IR) vacuums for $a \to 0$. Expanding (\ref{CurvedBPSEq1})- (\ref{CurvedBPSEq2}) around $\mathbf{\Phi}_{c}$, we then obtain the eigenvalues of the first order expansion matrix 
\begin{equation}
\lambda^{(0)}_{\mp} = \mp \frac{(D-1)}{\kappa^{2} }  g^{z\bar{z}}(\mathbf{\Phi}_{c}) \lambda^{\mathcal{W}}_{\pm}  = \frac{(D-1)}{\kappa^{2} } \left( \mp \kappa^{2} \mathcal{W}(\mathbf{\Phi}_{c}) -  2  g^{z\bar{z}}(\mathbf{\Phi}_{c}) |\partial_z^2 \mathcal{W}(\mathbf{\Phi}_{c})|  \right)\ . \label{EigenGFlat}
\end{equation}
According to Lyapunov theorem, a stable flow requires that all eigenvalues of (\ref{EigenGFlat}) have to be negative which follows \cite{GZA2007} 
\begin{equation}
|\partial^2_z\mathcal{W}\left(\mathbf{\Phi}_{c}\right)| > \frac{\kappa^{2}   }{2} g_{z\bar{z}}\left(\mathbf{\Phi}_{c}\right) \mathcal{W}\left(\mathbf{\Phi}_{c}\right) \ ,
\label{CurvedMinVflat}
\end{equation}
 flowing along local minima and the stable directions of saddles of the scalar potential (\ref{CurvedPotential}). In other words, the evolution of the complex scalars $(z, \bar{z})$ is stable on the walls  in the context of dynamical system. In particular, if the sliced geometry is flat Minkowskian, then we have a \textit{complete stable} model which means that it is stable both in the context of gravitational stability and in the context of dynamical system. Other models would be \textit{incomplete stable}.

In the case at hand,  (\ref{ScaleNearCP}) simplifies to 
\begin{equation}
a_{\mathrm{rf}}(u) = a_{0} \ e^{\mathcal{W}(\mathbf{\Phi}_{c})u} \ , \label{FlatScaleNearCP}
\end{equation}
which  shows that at  $u \rightarrow +\infty$, we have $a_{\mathrm{rf}} \rightarrow +\infty$ (the UV vacuums), while  at $u\rightarrow -\infty$, we have $a_{\mathrm{rf}}\rightarrow 0$ (the IR vacuums). The eigenvalues of first order expansion of the beta function (\ref{Betaeq}) are 
\begin{equation}
\tilde{\lambda}^{(0)}_{\pm} =  \frac{(D-1)}{\kappa^{2} } \frac{g^{z\bar{z}}(\mathbf{\Phi}_{c})}{{\mathcal{W}}(\mathbf{\Phi}_{c})}  \lambda^{\mathcal{W}}_{\pm}  = \frac{(D-1)}{\kappa^{2} } \left(  \kappa^{2}  \pm  2  \frac{g^{z\bar{z}}(\mathbf{\Phi}_{c})}{{\mathcal{W}}(\mathbf{\Phi}_{c})}   |\partial_z^2 \mathcal{W}(\mathbf{\Phi}_{c})|  \right)\ . \label{EigenBetaFlat}
\end{equation}
In the UV region, we require at least one of the eigenvalues (\ref{EigenBetaFlat}) to be positive. So, the RG flow is stable along $\tilde{\lambda}^{(0)}_+$, but it might be unstable in the direction of $\tilde{\lambda}^{(0)}_-$ if $\tilde{\lambda}^{(0)}_- \le 0$ which implies that the flow fails to depart the UV region. In conclusion, we have here all type of vacuums, namely local maxima (unstable), local minima (stable), saddles (stable), and bifurcation points  (undetermined). On the other hand, in the IR region the RG flow is stable along the direction of the negative eigenvalue of $\tilde{\lambda}^{(0)}_-$, but it might be failed to  approach vacuums if $\tilde{\lambda}^{(0)}_- \ge 0$. Thus, there are no local maxima and bifurcation points.\\
\indent In Einstein domain walls, the eigenvalues of the first order expansion of gradient flow equations (\ref{CurvedBPSEq1})- (\ref{CurvedBPSEq2}) are
\begin{equation}
\lambda^{(\Lambda_D)}_{\mp} = \mp \frac{(D-1)}{\kappa^{2} \gamma^2}  g^{z\bar{z}}(\mathbf{\Phi}_{c}) \lambda^{\mathcal{W}}_{\pm}  = \frac{(D-1)}{\kappa^{2}  \gamma^2 } \left( \mp \kappa^{2} \mathcal{W}(\mathbf{\Phi}_{c}) -  2  g^{z\bar{z}}(\mathbf{\Phi}_{c}) |\partial_z^2 \mathcal{W}(\mathbf{\Phi}_{c})|  \right) \ , \label{EigenGCurved}
\end{equation}
whereas the eigenvalues of first order expansion of the beta function (\ref{Betaeq}) have the form
\begin{equation}
\tilde{\lambda}^{(\Lambda_D)}_{\pm} =  \frac{(D-1)}{\kappa^{2} \gamma^2} \frac{g^{z\bar{z}}(\mathbf{\Phi}_{c})}{{\mathcal{W}}(\mathbf{\Phi}_{c})}  \lambda^{\mathcal{W}}_{\pm}  = \frac{(D-1)}{\kappa^{2} \gamma^2} \left(  \kappa^{2}  \pm  2  \frac{g^{z\bar{z}}(\mathbf{\Phi}_{c})}{{\mathcal{W}}(\mathbf{\Phi}_{c})}   |\partial_z^2 \mathcal{W}(\mathbf{\Phi}_{c})|  \right)\ . \label{EigenBetaCurved}
\end{equation}
Comparing (\ref{EigenGCurved}) - (\ref{EigenBetaCurved}) with (\ref{EigenGFlat}) and (\ref{EigenBetaFlat}), it is  easily to see that we will get the same results as the previous Ricci-flat case.  These results, however, need the analysis on the warped factor (\ref{ScaleNearCP}) in order to have a well-defined field theory on the boundaries. First, if we want the UV vacuums ($a \rightarrow +\infty$) and the IR vacuums (($a \rightarrow 0$) ) in the limit of  $u \rightarrow \pm \infty$, then  the sliced spacetime should be a weak Ricci-flat with small cosmological constant, namely $\Lambda_{D} \approx 0$, or  we have a case of $\Lambda_{D} \ll  \mathcal{W}^2(\mathbf{\Phi}_{c})$. Second, we modify  the theory such that a well-defined field theory could only be defined on the interval $- |\Lambda_{D}|^{-1} \le u \le |\Lambda_{D}|^{-1} $. The UV vacuums can be introduced in the limit of $u \to |\Lambda_{D}|^{-1}$, while the IR vacuums are in the limit of $u \to - |\Lambda_{D}|^{-1}$ where $a(|\Lambda_{D}|^{-1}) > a(-|\Lambda_{D}|^{-1})$. Note that our analysis fails for the case of $|\Lambda_{D}| \to +\infty$.

\section{Simple Models}

\subsection{Flat K\"ahler Models}
In this subsection we apply the previous general setup to   a simple  model  where the  K\"ahler space is flat with potential $K(z,\bar{z}) = z\bar{z}$  and the superpotential has the quadratic form, namely $W(z) = \alpha_0 + \alpha_1 z + \alpha_2 z^2$ where $\alpha_0$, $\alpha_1$, $\alpha_2$ are real constants. Here, we assume the origin to be a critical point of this model which implies that  $K_z(0) = 0$. Then,   (\ref{eigenW}) and (\ref{CurvedEigenV}) simplify to
\begin{eqnarray}
\lambda^{\mathcal{W}}_{\pm}  &=&  \kappa^{2} |\alpha_0 | \pm  2 |\alpha_2| \ , \\
\lambda^V_{\pm} & = & \frac{2}{\kappa^{4}\gamma^{2}}(D-1)^2  |\alpha_2|^2 + (D-1)\left[\frac{D-1}{2\gamma^{2}}-D\right] |\alpha_0|^2 \nonumber\\
& &\pm \frac{2}{\kappa^{2}}(D-1)\left[D-\frac{D-1}{\gamma^{2}}\right]  |\alpha_0|  |\alpha_2| \ ,\\
\end{eqnarray}
and the scalar curvature of the ambient spacetime equals $- D (D+1)  |\alpha_0| $.

For $\alpha_0 = 0$, we have a complete stable Minkowskian ground state with $\Lambda_D = 0$, while an AdS ground state requires $\alpha_0 \ne 0$ for any $\Lambda_D$. In the latter case, we have an AdS vacuum for $\alpha_2 = 0$ which is  unstable in the context dynamical system, but for $\Lambda_D = 0$ it is easy to check that this vacuum is gravitationally stable. In other words, this vacuum is incomplete stable. The anaysis on RG flow shows that this vacuum exists only in the UV region.\\
\indent It is worth mentioning that in four dimensions (\textit{i.e.} $D=3$) $N=1$ supersymmetry further demands that we have  three dimensional Minkowskian  spacetimes as studied in \cite{GZA2007}. On the other hand, our setting in this paper does not cover the  curved AdS sliced BPS domain walls \cite{GZ2009} since these cases do not satisfy the Einstein field equations and the scalar equations of motions. In other words, these only exist in the off-shell formulation.

\subsection{$5d ~ N=2$ Supergravity Coupled To A Hypermultiplet}

In this subsection we show that our general setting in the previous section is related to a case  of BPS domain walls in five dimensional $N=2$ supergravity coupled only to a hypermultiplet where the hyperscalars span a four dimensional  quaternionic K\"ahler manifold, which is Einstein and has also a selfdual property that is  selfdual  Weyl tensor. This class of four-manifolds has been constructed in \cite{Calderbank:2001dg} which covers homogeneous and inhomogeneous quaternionic K\"ahler geometries.\\
%
%
\indent  First of all, we write down the metric of four dimensional hyperscalar manifolds as \cite{Calderbank:2001dg}
\begin{equation}
ds^2 = \Big[
\frac{1}{4{\rho}^2} - \frac{(F^2_{\rho}+F^2_{\eta})}{F^2}\Big]
\big( d{\rho}^2 + d{\eta}^2 \big)
 +\frac{\left[ \big(F - 2\rho F_{\rho})\alpha - 2\rho F_{\eta}\beta \right]^2
+\left[ -2\rho F_{\eta}\alpha+(F + 2\rho F_{\rho})\beta \right]^2}
{F^2 \big( F^2 - 4{\rho}^2(F^2_{\rho} + F^2_{\eta}) \big)},
\label{metricTSDE}
\end{equation}
where  the hyperscalars are $ (\rho, \eta, \phi, \varphi)$, whereas   $\alpha=\sqrt{\rho}\,d\phi$, $\beta=(d\varphi + \eta\, d\phi)$, and $(\phi,\varphi)$ are periodic coordinates. The function $F(\rho,\eta)$ satisfies the Laplace equation in two dimensional hyperbolic space
\begin{equation}\label{diff:eps}
\rho^2 (F_{\rho\rho}+F_{\eta\eta})=\frac{3}{4}F,
\end{equation}
with $F_{\rho\rho} \equiv \frac{{\partial}^2{F}}{\partial{\rho}^2}$ and
$F_{\eta\eta} \equiv \frac{{\partial}^2{F}}{\partial{\eta}^2}$.  Furthermore, it has positive scalar curvature if $F$ satisfies
$F^2 > 4{\rho}^2(F^2_{\rho}+F^2_{\eta})$ and negative if $F^2 <
4{\rho}^2(F^2_{\rho}+F^2_{\eta})$. Clearly that the metric (\ref{metricTSDE}) has $T^2$ isometry along the periodic coordinates $(\phi, \varphi)$.  \\
\indent Next, let us first choose the isometry of the metric (\ref{metricTSDE}) generated by
\begin{equation}
k_0 = \partial_{\phi}- g\, \partial_{\varphi} \ , \label{kisometry}
\end{equation}
where we have normalized $k_0$ such that the prefactor of $\partial_{\phi}$ equals one and $g$ is a real constant. We can then obtain the triplet of prepotential
\begin{equation}
P^1 =0\:, \quad P^2 = \frac{1}{2} \frac{\sqrt{\rho}}{F}\:, \quad P^3 = \frac{1}{2} \frac{\eta-g}{\sqrt{\rho}\,F} \label{prepTSDE}\;,
\end{equation}
which can be used to compute the real fake superpotential \cite{Anguelova:2002gd}
\begin{equation}
 {\mathcal W} = \sqrt{\frac{1}{6F^2}\left[ \rho+ \frac{(\eta-g)^2}{\rho}\right]} \ . \label{realsuperpot}
\end{equation}
Defining the complex coordinate $z \equiv \rho + {\mathrm{i}} \eta$, it is straightforward to see that this two dimensional metric is K\"ahler with  K\"ahler potential 
\begin{equation}
 K \equiv {\mathrm{ln}}(z + \bar{z}) + 2 \ {\mathrm{ln}} F  \ , \label{Kpot}
\end{equation}
and the holomorphic superpotential has the form
\begin{equation}\label{holomorW}
W(z) = \frac{1}{\sqrt{3}} ( z - \mathrm{i} g) \ .
\end{equation}
Then, the gradient flow equations in (\ref{CurvedBPSEq1}) can be rewritten as
\begin{equation}
z' = \mp \ \frac{3}{\kappa^2 \gamma} g^{z\bar{z}}  \partial_{\bar{z}} {\mathcal W}  \ ,\label{susyflow1}
\end{equation}
together with its complex conjugate. The quantity $g^{z\bar{z}}$ is the inverse of the K \"ahler metric $g_{z\bar{z}}  \equiv \partial_z \partial_{\bar{z}} K$ where $K$ is the K\"ahler potential introduced in (\ref{Kpot}). It is worth mentioning that since the metric (\ref{metricTSDE}) is definite positive for both positive and negative scalar curvature cases, then the metric $g_{z\bar{z}} $ should also be definite positive.\\
\indent The scalar potential (\ref{CurvedPotential})  has simply the form
\begin{equation}
\kappa^2 V = \frac{9}{2 \kappa^2 \gamma^2} g^{z\bar{z}}   \partial_{z}{\mathcal W}  \partial_{\bar{z}} {\mathcal W} - 6 {\mathcal W}^2 \ , \label{V3}
\end{equation}
after employing the method discussed in \cite{CDL2002}, while the beta function (\ref{Betaeq}) has a complex value
\begin{equation}
\beta^z  = - \frac{3}{\kappa^2 \gamma^2} g^{z\bar{z}} \frac{\partial_{\bar{z}} {\mathcal W} }{{\mathcal W}}  \  . \label{betaqX1}
\end{equation}
\indent Note that we have a class of quaternionic manifolds in which it does not have any BPS domain walls in this setup. Such a case exists if
\begin{equation}
F = \sqrt{ \rho+ \frac{(\eta-g)^2}{\rho}}  \ .
\end{equation}
The superpotential (\ref{realsuperpot})  becomes a constant and the solution is anti-de Sitter spacetime.  In other words, this corresponds to the trivial K\"ahler potential which is excluded  in our case.

\section*{Acknowledgements}

We thank anonymous referee for the suggestion of this paper. Our research is supported by Riset KK ITB 2015  and Riset Desentralisasi DIKTI-ITB 2015.

\end{document}